\documentclass[11pt, a4paper, Physsubmission, Phys]{SciPost} 

\hypersetup{
    colorlinks,
    linkcolor={red!50!black},
    citecolor={blue!50!black},
    urlcolor={blue!80!black}
}

\usepackage[bitstream-charter]{mathdesign}
\DeclareSymbolFont{usualmathcal}{OMS}{cmsy}{m}{n}
\DeclareSymbolFontAlphabet{\mathcal}{usualmathcal}

\begin{document}

\begin{center}{\Large \textbf{
Muon puzzle in inclined muon bundles\\ detected by NEVOD-DECOR\\
}}\end{center}

\begin{center}
A.G. Bogdanov\textsuperscript{1$\star$},
N.S. Barbashina\textsuperscript{1}, 
S.S. Khokhlov\textsuperscript{1}, 
V.V. Kindin\textsuperscript{1}, 
R.P. Kokoulin\textsuperscript{1}, \\
K.G. Kompaniets\textsuperscript{1}, 
A.Yu. Konovalova\textsuperscript{1}, 
G. Mannocchi\textsuperscript{2}, 
A.A. Petrukhin\textsuperscript{1}, 
V.V. Shutenko\textsuperscript{1}, \\
G. Trinchero\textsuperscript{2},
V.S. Vorobev\textsuperscript{1}, 
I.I. Yashin\textsuperscript{1}, 
E.A. Yurina\textsuperscript{1}, 
and
E.A. Zadeba\textsuperscript{1}
\end{center}

\begin{center}
\mbox{{\bf 1} National Research Nuclear University MEPhI (Moscow Engineering Physics Institute), Russia}
\\
{\bf 2} Osservatorio Astrofisico di Torino -- INAF, Italy
\\
* agbogdanov@mephi.ru
\end{center}

\begin{center}
\today
\end{center}


\definecolor{palegray}{gray}{0.95}
\begin{center}
\colorbox{palegray}{
  \begin{tabular}{rr}
  \begin{minipage}{0.1\textwidth}
    \includegraphics[width=30mm]{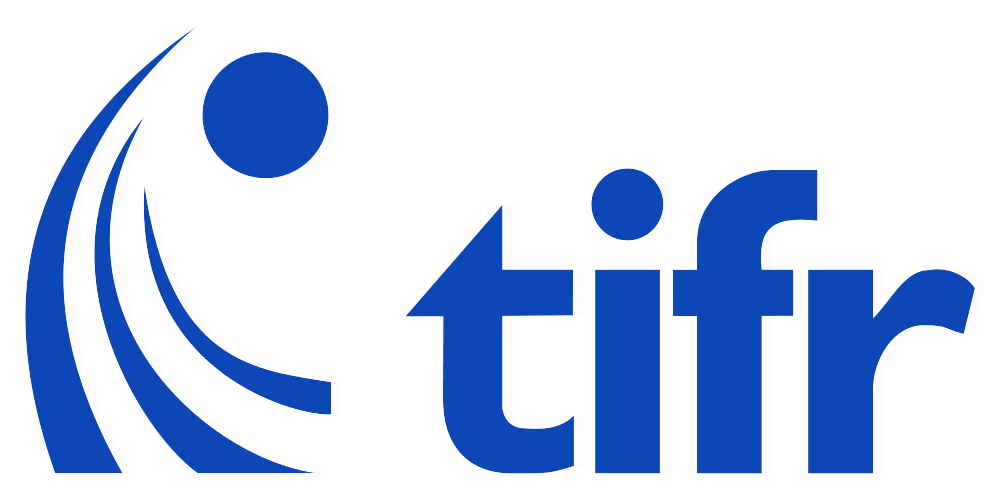}
  \end{minipage}
  &
  \begin{minipage}{0.85\textwidth}
    \begin{center}
    {\it 21st International Symposium on Very High Energy Cosmic Ray Interactions (ISVHECRI 2022)}\\
    {\it Online, 23-27 May 2022} \\
    \doi{10.21468/SciPostPhysProc.?}\\
    \end{center}
  \end{minipage}
\end{tabular}
}
\end{center}

\section*{Abstract}
{\bf
The data of cosmic ray NEVOD-DECOR experiment on the investigation of inclined muon bundles for a long time period (May 2012 -- March 2021) are presented. The analysis showed that the observed intensity of muon bundles at primary cosmic ray energies of about 1 EeV and higher can be compatible with the expectation in frame of widely used hadron interaction models only under the assumption of an extremely heavy mass composition. This conclusion is consistent with data of several experiments on investigations of muon content in air showers, but contradicts the available fluorescence data on $X_{\text{max}}$ which favor a light mass composition at these energies. In order to clarify the nature of  the ``muon puzzle'', investigations of the muon bundle energy deposit in the detector material were carried out. For the first time, experimental estimates of the average energy of muons in the bundles of inclined air showers initiated by primary particles with energies from 10 to 1000 PeV have been obtained.
}

\vspace{10pt}
\noindent\rule{\textwidth}{1pt}
\tableofcontents\thispagestyle{fancy}
\noindent\rule{\textwidth}{1pt}
\vspace{10pt}

\section{Introduction}
\label{sec:intro}
The muon component of extensive air showers (EAS) is formed mainly as a result of the decays of pions and kaons generated in hadronic interactions in the atmosphere, and is often used to estimate the mass composition of primary cosmic rays (PCR) and to verify models of hadron interactions at high energies. Of particular interest are events with muon bundles, which represent a simultaneous (within tens of nanoseconds) passage through the detector of several penetrating particles with almost parallel tracks.

To date, NEVOD-DECOR is the only experiment in which long-term systematic studies of muon bundles in a wide range of zenith angles are conducted. This article provides a brief overview of the results of these studies obtained so far.

\section{NEVOD-DECOR setup and experimental data}
\label{sec:experiment}

The NEVOD-DECOR setup is located at MEPhI and includes a Cherenkov water calorimeter NEVOD \cite{1} with a volume of about 2000 m$^3$ and a coordinate-tracking detector DECOR \cite{2} with an area of about 70 m$^2$, united by a trigger system.

The Cherenkov water detector NEVOD consists of 91 quasi-spherical measuring modules (QSMs) located at the nodes of the volume-centered spatial lattice. In fact, the lattice is formed by 25 vertical strings of 3 or 4 QSMs. Each QSM consists of 6 FEU-200 photomultipliers with a flat photocathode of 15 cm in diameter, oriented along the axes of the orthogonal coordinate system. Such design provides almost the same detection efficiency of Cherenkov radiation coming from any direction. A wide dynamic range of the measurements of each PMT (1 -- 10$^5$ photoelectrons) is due to two-dynode signal readout. This allows calorimetric studies, in particular, measurements of the energy deposit of muon bundles.

The coordinate-tracking detector DECOR consists of 8 vertical supermodules (SMs) located in the galleries of the experimental building on three sides of the water volume of the NEVOD detector. The effective area of one SM is 3.1$\times$2.7 m$^2$. Each SM represents 8 parallel planes consisting of plastic chambers of gas-discharge streamer tubes. The planes are equipped with a system of external strips for signal readout in two coordinates (along and across the chambers). Accuracy of localization of tracks of charged particles in one SM is better than 1 cm, and the angular accuracy of the reconstruction of tracks crossing the SM is better than 1$^{\circ}$.

In the present work, we used experimental data on muon bundles accumulated over a long period from May 2012 to March 2021 (``live'' observation time is equal to 58.3 thousand hours). About 99.6 thousand events with muon multiplicity $m \ge 5$ and zenith angles $\theta \ge 55^{\circ}$ were selected; in addition 30.4 thousand events with smaller zenith angles from 40 to 55$^{\circ}$ (``live'' time is 6.3 thousand hours) were sampled. In order to improve the identification of muon tracks, the events were selected in two 60$^{\circ}$-wide azimuth angle sectors where 6 of the 8 DECOR SMs are shielded by the NEVOD detector water tank. At that, the average threshold energy of muons is about 2 GeV.

The method of identifying muon bundles of atmospheric origin in a coordinate detector is based on the parallelism of particle tracks (within 5$^{\circ}$ cone) recorded by the setup. The event selection procedure consists of several stages: hardware level (3-fold coincidence of the signals from different SMs within the time gate of 250 ns); program reconstruction and selection; final classification of events and counting of tracks by several operators.

An example of the spatial reconstruction of the muon bundle ($m$ = 8 and $\theta$ = 57$^{\circ}$) registered in the NEVOD-DECOR setup, as well as the main parts of the experimental complex are shown in Figure 1. Thin lines show the reconstruction of muon tracks according to the DECOR data, the small circles show the hit PMTs in the NEVOD calorimeter (colors reflect signal amplitudes), the large rectangles show the SMs of the detector DECOR. Multimuon events have a bright signature in the coordinate detector, and their interpretation is almost unambiguous. Obviously, for each such event, it is possible to find the number of tracks and determine the direction. 

\begin{figure}[h]
\centering
\includegraphics[width=.95\linewidth]{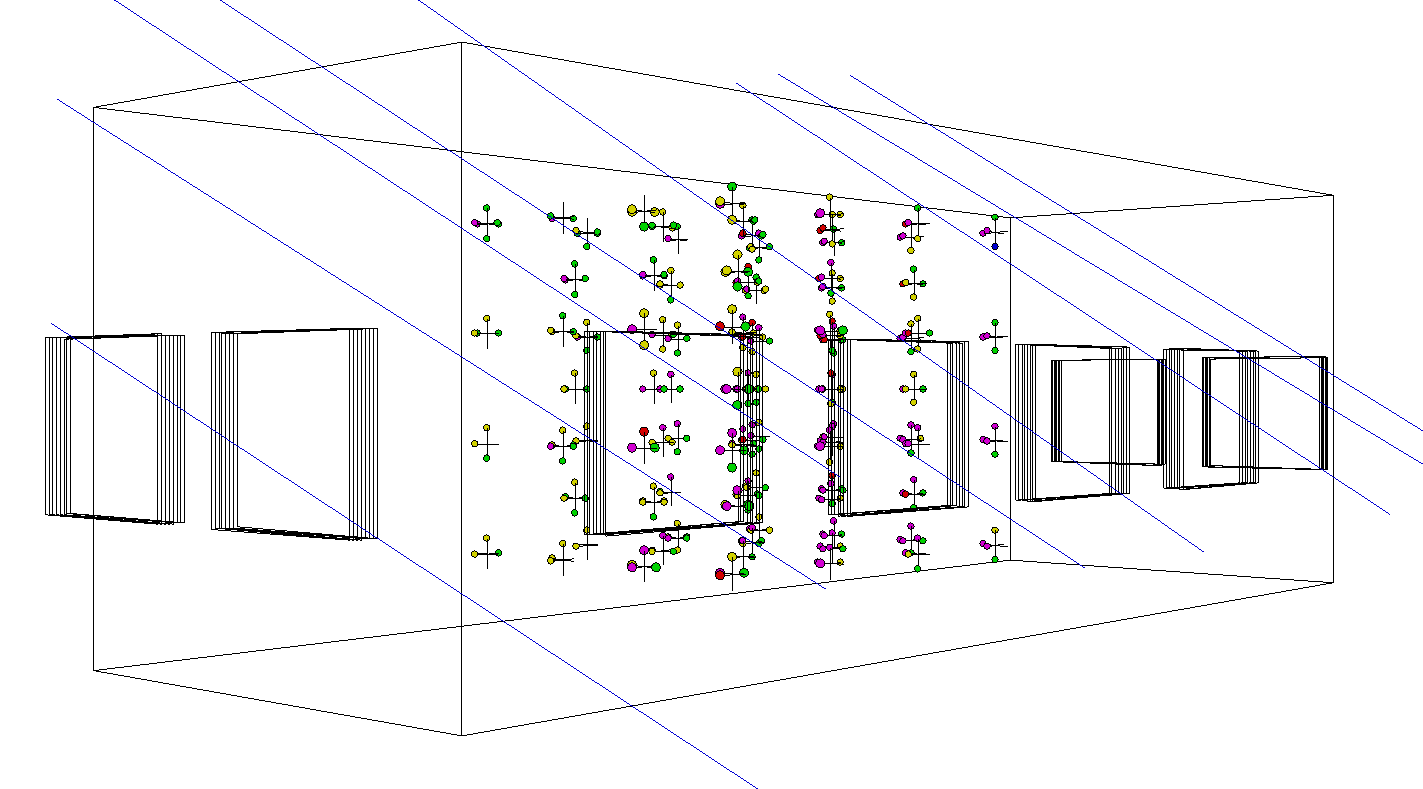}
\vspace{-0.25cm}
\caption{An example of geometrical reconstruction of muon bundle event detected by the NEVOD-DECOR setup}
\label{event}
\end{figure}

During the whole time of the experiment, an express analysis of the NEVOD and DECOR data is carried out in order to quickly identify and eliminate equipment faults and to prepare experimental data for further physical and methodical processing.

\section{Local muon density spectra}
\label{sec:LMDS}

A new approach to the study of the EAS, the method of local muon density spectra (LMDS), was developed for the physical analysis of the data of the first experiment on the study of muon bundles conducted at detector DECOR in 2002-2007  \cite{3}. The typical size of the EAS muon component ($\sim$ km) significantly exceeds the size of the NEVOD-DECOR setup (tens of meters), so the detector can be considered as a point-like one. In an individual event with a muon bundle the local muon density $D$ at the observation point is estimated, in the first approximation: $D \sim m/S_{\text{det}}$, where $m$ is the multiplicity of muons in the bundle (the number of muons that hit the detector), and $S_{\text{det}}$ is the area of the DECOR detector for a given direction. The distribution of events in the local muon density estimate $D$ forms the LMDS. The spectrum of events by local muon density has a nearly power-law form, with a slope $\beta$ slightly steeper than the slope of the PCR spectrum.

It is important that, at the same muon density, different zenith angles correspond to substantially (by orders of magnitude) different characteristic energies of primary particles contributing to the selected events. The fact is that with an increase in the zenith angle, the spread of muons in bundles increases, mainly due to transverse momenta during the generation and decay of mother hadrons, deviation in the Earth's magnetic field and multiple scattering, since muons travel a longer path in the atmosphere. Thus, measurements of the LMDS at different zenith angles make it possible to study a wide energy range of PCRs by means of a relatively small setup. In this case, the event collection area is determined not by the size of the detector, but by the cross section of the air shower, which in the muon component near the horizon reaches several km$^2$, that is sufficient to reach energies of 10$^{18}$ eV and even higher.

The local muon density spectra are used as a frontier between the experiment and calculations. The main stages of data analysis on the local muon density are described in detail in \cite{3,4,5}.

The experimental spectra are obtained by means of reconstructing $dF(D,\theta)/dD$ (in a detector-independent form) from the experimental distributions of event characteristics $N(m,\theta,\phi)$. The reconstruction procedure takes into account geometric factors, Poisson fluctuations in the number of muons that hit the detector, the detection efficiency of streamer tube chambers, trigger and event selection conditions, etc.

The expected spectra $dF(D,\theta)/dD$ are obtained as a result of convolution of the lateral distribution functions (LDFs) of muons with certain models of the PCR spectrum. Two-dimensional LDFs are calculated on the basis of EAS simulations using the CORSIKA program \cite{6} for a set of fixed zenith angles and PCR energies, also taking into account such factors as the threshold energy of secondary particles, the altitude of the observation level, the atmosphere, and the Earth’s magnetic field. Naturally, the shape of the LDF depends on the chosen model of hadron interactions and assumptions about the PCR mass composition.

Comparison of experimental and calculated LMDS for 9 intervals of zenith angles is shown in Figure 2. The NEVOD-DECOR data are marked by symbols. The solid, dashed, and dash-dotted curves represent the results of calculations based on the CORSIKA (v. 7.69) program for the hadronic interaction models tuned according to the LHC data: QGSJET-II-04, SIBYLL-2.3c, and EPOS-LHC, respectively. Two limiting cases of the PCR mass composition were used in the simulations: only protons (p, three lower curves for each zenith angle) and only iron nuclei (Fe, upper curves). The absolute intensity increases for showers initiated by heavier nuclei, because at a fixed nucleus energy $E_{0}$, the muon density rises with increasing atomic mass. As a model of the PCR energy spectrum, the approximation of the data of various experiments from the PDG review \cite{7} proposed by us in \cite{4} was used. The arrows in the figure indicate the typical (mean logarithmic) PCR energies, which give the main contribution to the events sampled by the muon density. In this case, the PCR energy distributions are relatively narrow ($\sigma_{\text{lgE}}$ $\approx$ 0.4 for $D$ = 0.2, which corresponds to 5 -- 7 muons that passed through the detector DECOR). This is due to a rapid decrease in the cosmic ray flux with increasing energy, although the common integral is given by air showers from PCRs with different energies, registered at different (random) distances from the EAS axis.

\begin{figure}[h]
\centering
\includegraphics[width=1.\linewidth]{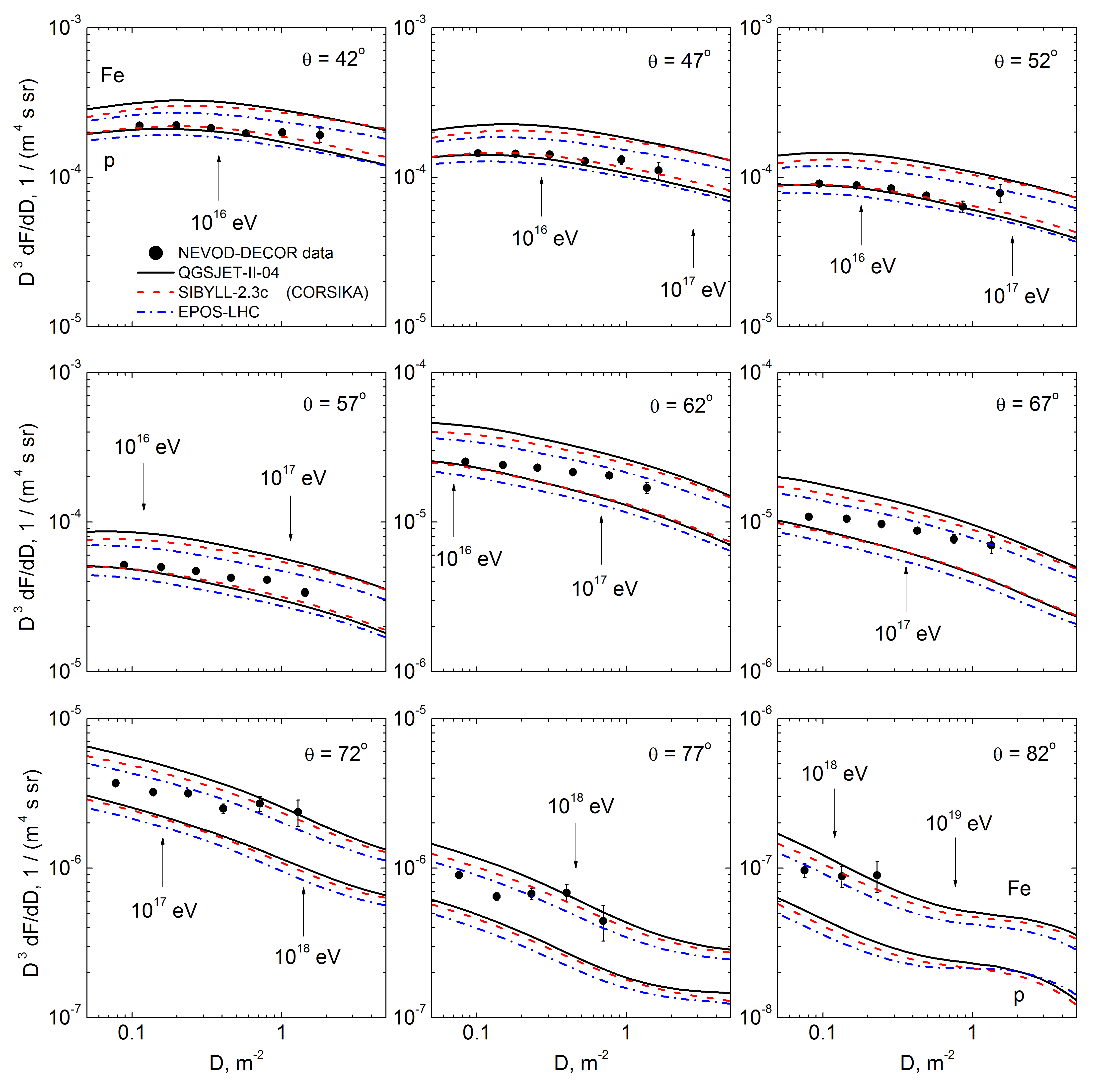}
\vspace{-0.5cm}
\caption{Differential local muon density spectra for different zenith angles}
\label{lmds}
\end{figure}

It can be seen from the figure that at moderate zenith angles (42 -- 52$^{\circ}$), which correspond to PCR energies $\sim$ 10$^{16}$ eV, the experimental data are close to the results of calculations for a light mass composition. Further, at higher energies, there is a relative increase of the intensity of muon bundles, which, in principle, can be interpreted as a trend toward heavier PCR mass composition. At large zenith angles (77 -- 82$^{\circ}$), corresponding to PCR energies $\sim$ 10$^{18}$ eV, the NEVOD-DECOR data are compatible with calculations only under the assumption of an extremely heavy mass composition of PCRs (iron nuclei), if we remain within the framework of existing hadron interaction models. However, this contradicts the data on the measurements of the EAS development maximum $X_{\text{max}}$ by the fluorescence method (mainly the electron-photon component) \cite{8,9,10}, which indicate a light mass composition of PCRs at such energies.

For the comparison of LMDS at different zenith angles, we used the $z$-parameter, which was proposed by the Working group on Hadronic Interactions and Shower Physics (WHISP) \cite{11} to compare data of various experiments on the study of EAS muon component, including NEVOD-DECOR:
\begin{equation}
z = \left( \text{ln}\,N_{\mu}^{\text{obs}} - \text{ln}\,N_{\mu}^{\text{p sim}} \right) \Big/ \left( \text{ln}\,N_{\mu}^{\text{Fe sim}} - \text{ln}\,N_{\mu}^{\text{p sim}} \right),
\end{equation}
where $N_{\mu}^{\text{obs}}$ is the observed value (muon density, number of muons, intensity of muon bundles, etc.), and $N_{\mu}^{\text{p sim}}$ and $N_{\mu}^{\text{Fe sim}}$ are the calculated estimates of this value for EASs formed by primary protons and iron nuclei, then $z$ = 0 means that PCRs consist only of protons, and $z$ = 1 corresponds to iron nuclei.

The dependences of the $z$-parameter on the PCR energy for 9 intervals of zenith angles and three modern hadron interaction models: QGSJET-II-04, SIBYLL-2.3c, and EPOS-LHC are shown in Figure 3,a-c. In this case, the same model of the PCR energy spectrum was used to construct the LMDS \cite{4}. It follows from the figures that the values of the $z$-parameter obtained in different ranges of zenith angles from 40 to 90$^{\circ}$ overlap and are in good agreement with each other within the errors. At the same time, steady trend of growth of $z$-parameter at the energies above 10$^{17}$ eV is observed. Comparison of the figures 3,a-c shows the influence of the choice of models of hadronic interactions on this dependence. Thus, for the EPOS-LHC model in comparison with the QGSJET-II-04 and SIBYLL-2.3c models, the $z$-parameter is shifted towards a heavier mass composition over the entire PCR energy range.

\begin{figure}[h]
\begin{minipage}[h]{0.49\linewidth}
\center{\includegraphics[width=1.\linewidth]{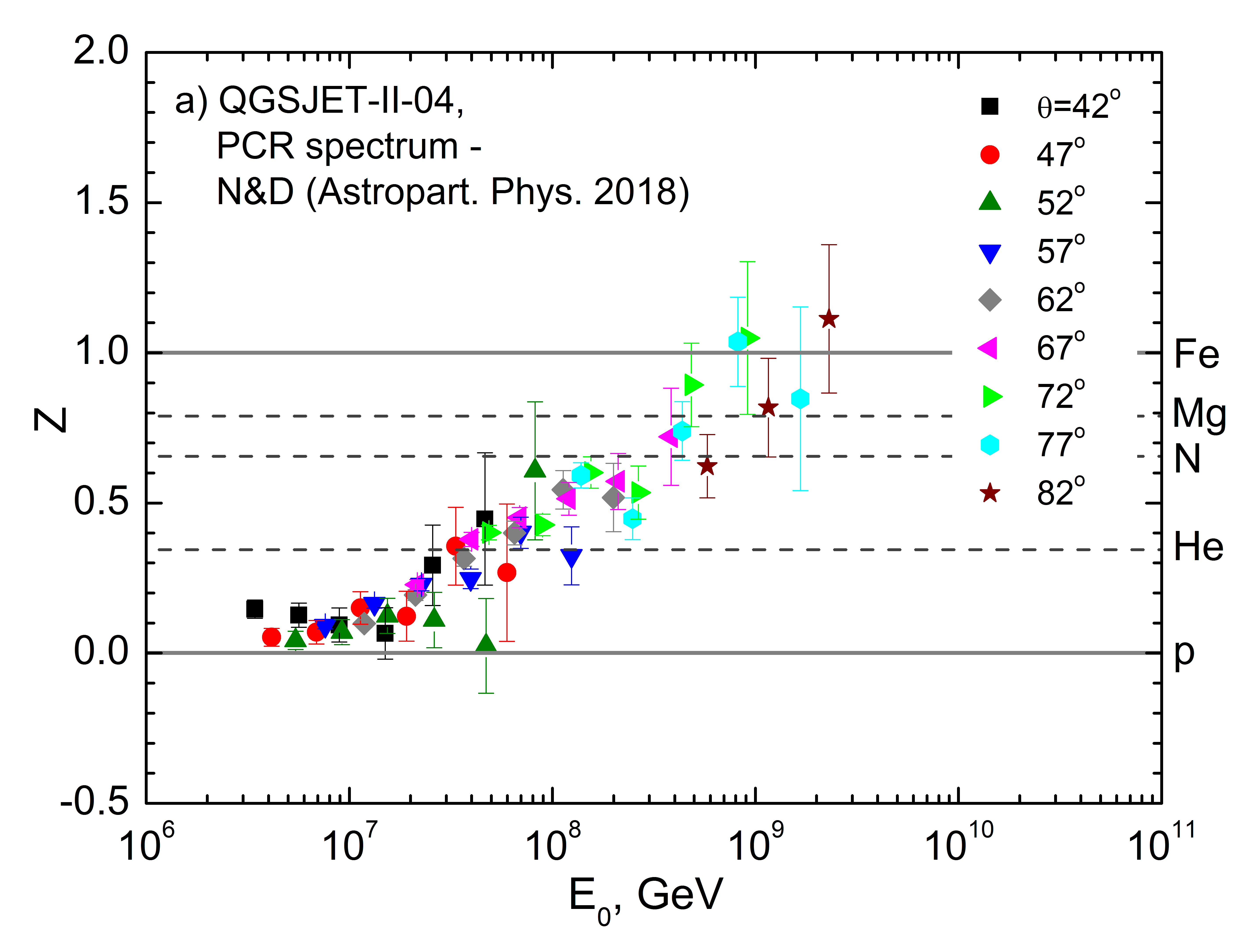}}
\end{minipage}
\hfill
\begin{minipage}[h]{0.49\linewidth}
\center{\includegraphics[width=1.\linewidth]{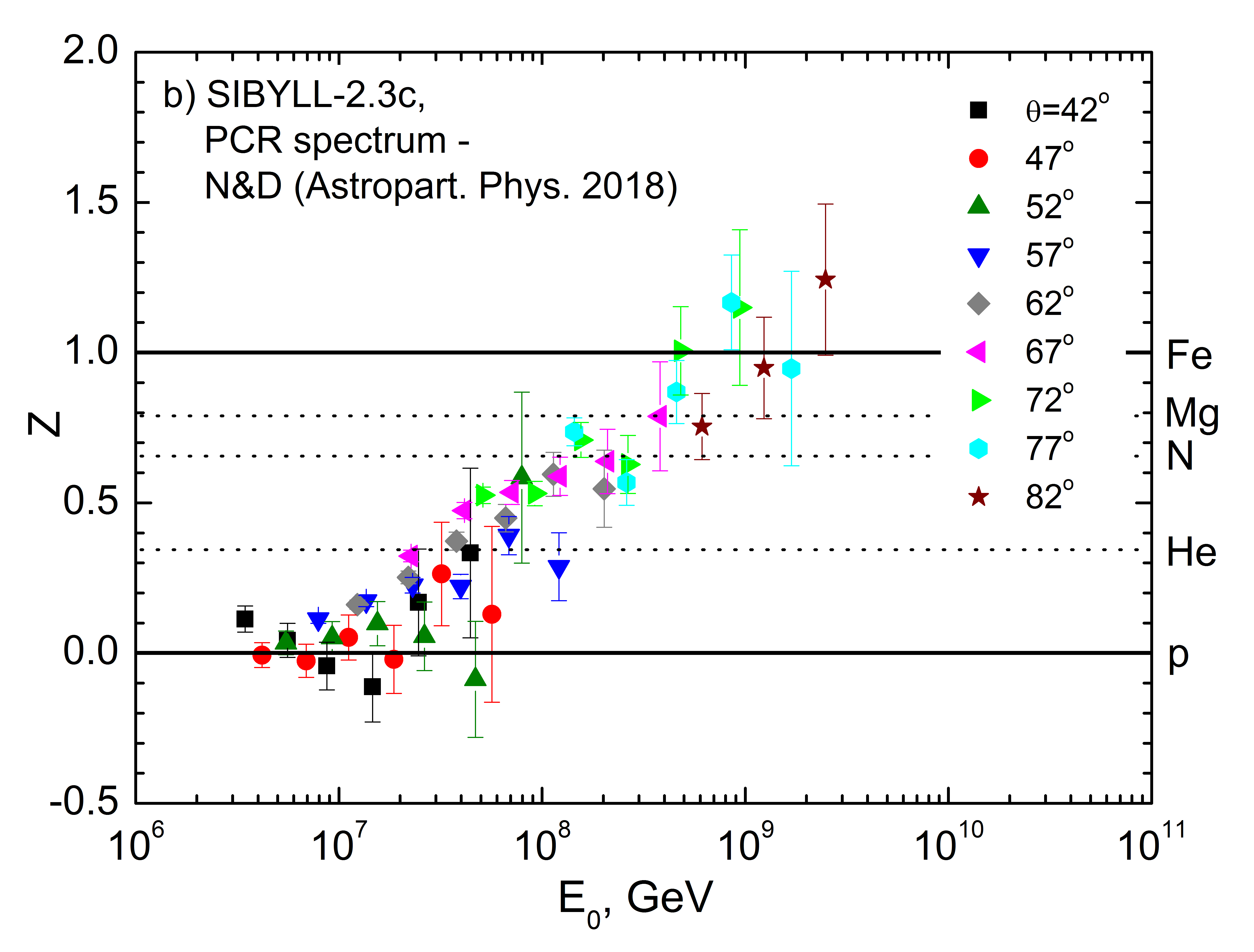}}
\end{minipage}
\begin{minipage}[h]{0.49\linewidth}
\center{\includegraphics[width=1.\linewidth]{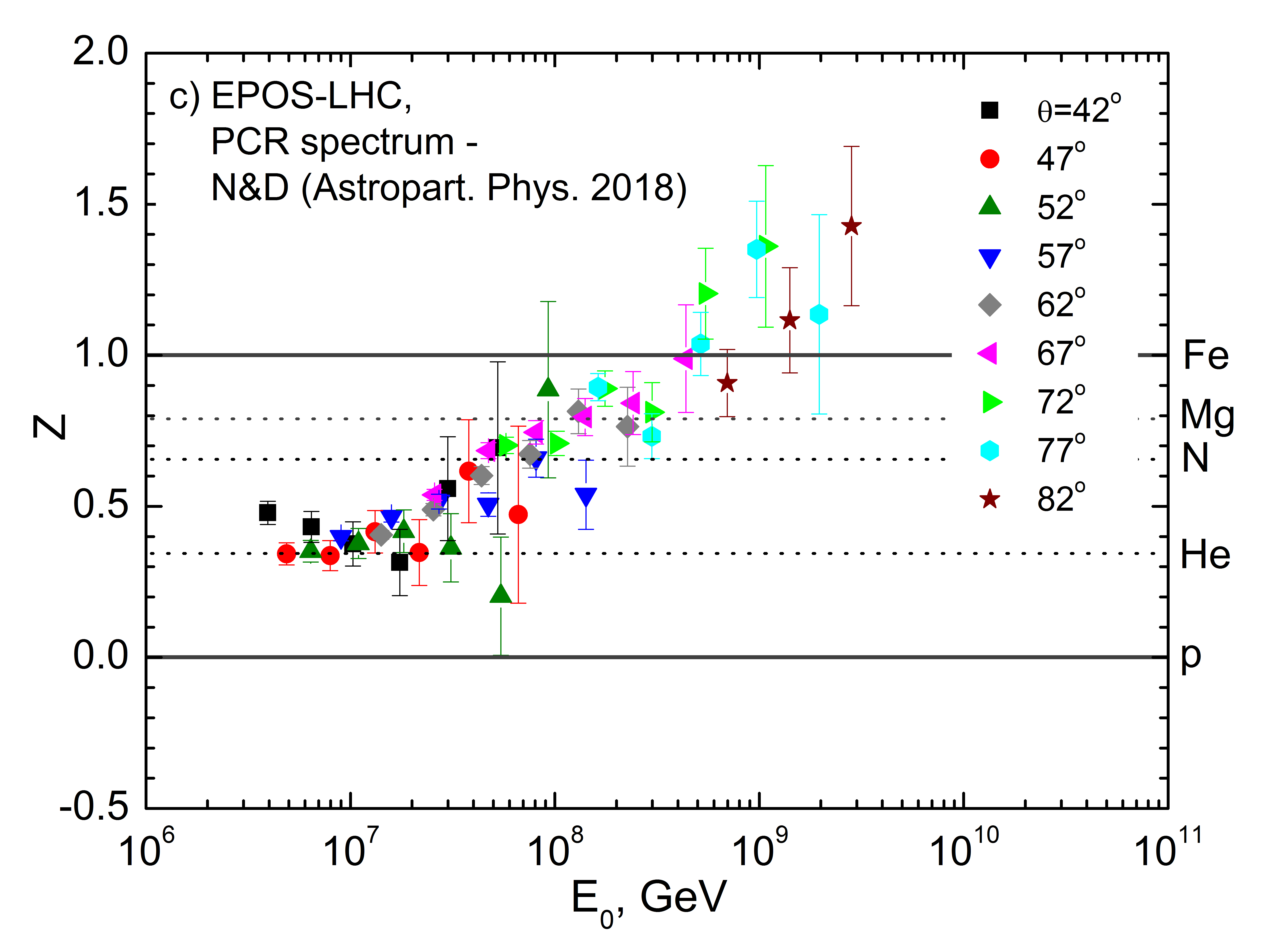}}
\end{minipage}
\hfill
\begin{minipage}[h]{0.49\linewidth}
\center{\includegraphics[width=1.\linewidth]{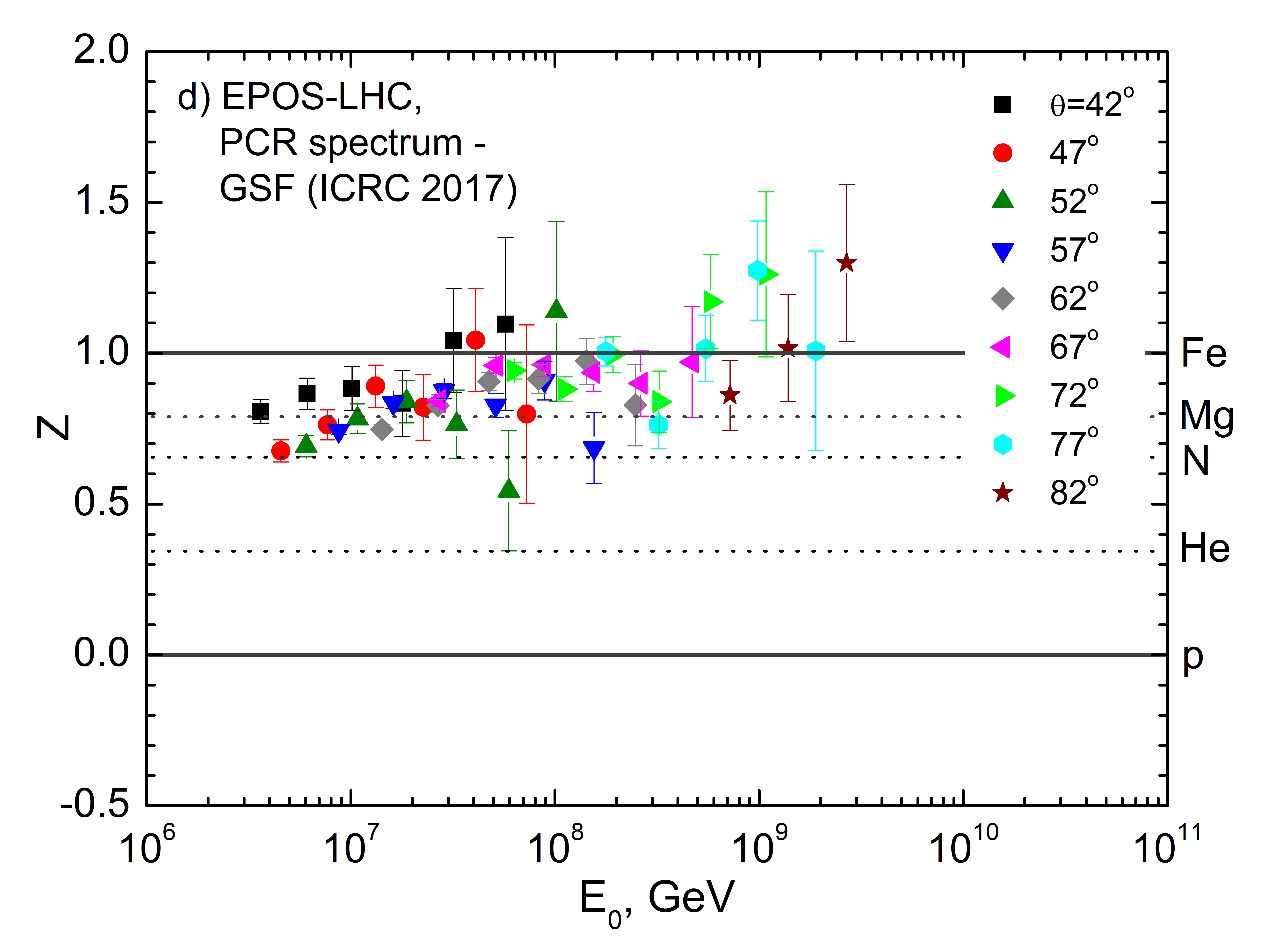}}
\end{minipage}
\caption{Comparison of LMDS at different zenith angles in terms of $z$-scale (a,b,c -- effect of the hadron interaction model; c,d -- influence of the assumed model of primary cosmic ray energy spectrum)}
\label{ris:zmodels}
\end{figure}

As mentioned above, in order to obtain the calculated LMDS, it is necessary to make an assumption about the shape of the PCR energy spectrum. In this regard, it is interesting to see how different versions of the spectrum affect the dependence of the $z$-parameter on the PCR energy. For comparison with our approximation from \cite{4}, we chose the GSF (Global Spline Fit) spectrum model proposed in \cite{12}, which is widely used by the WHISP group. It is a parametrization of data from various experiments, including direct observations, on the PCR flux and mass composition. 

A comparison of the dependences of the $z$-parameter on the PCR energy for these two variants of the spectrum and 9 intervals of zenith angles is shown in Figure 3c,d, in both cases the same model of hadronic interactions EPOS-LHC was used. Obviously, the choice of the PCR energy spectrum model affects both the absolute values of the $z$-parameter and its energy dependence (the slope changes). Both PCR spectra are shown in Figure 4. In general, their difference is not too large, but still quite noticeable (about 20\%) in the energy range from 10 to 100 PeV.

\begin{figure}[h]
\centering
\includegraphics[width=.5\linewidth]{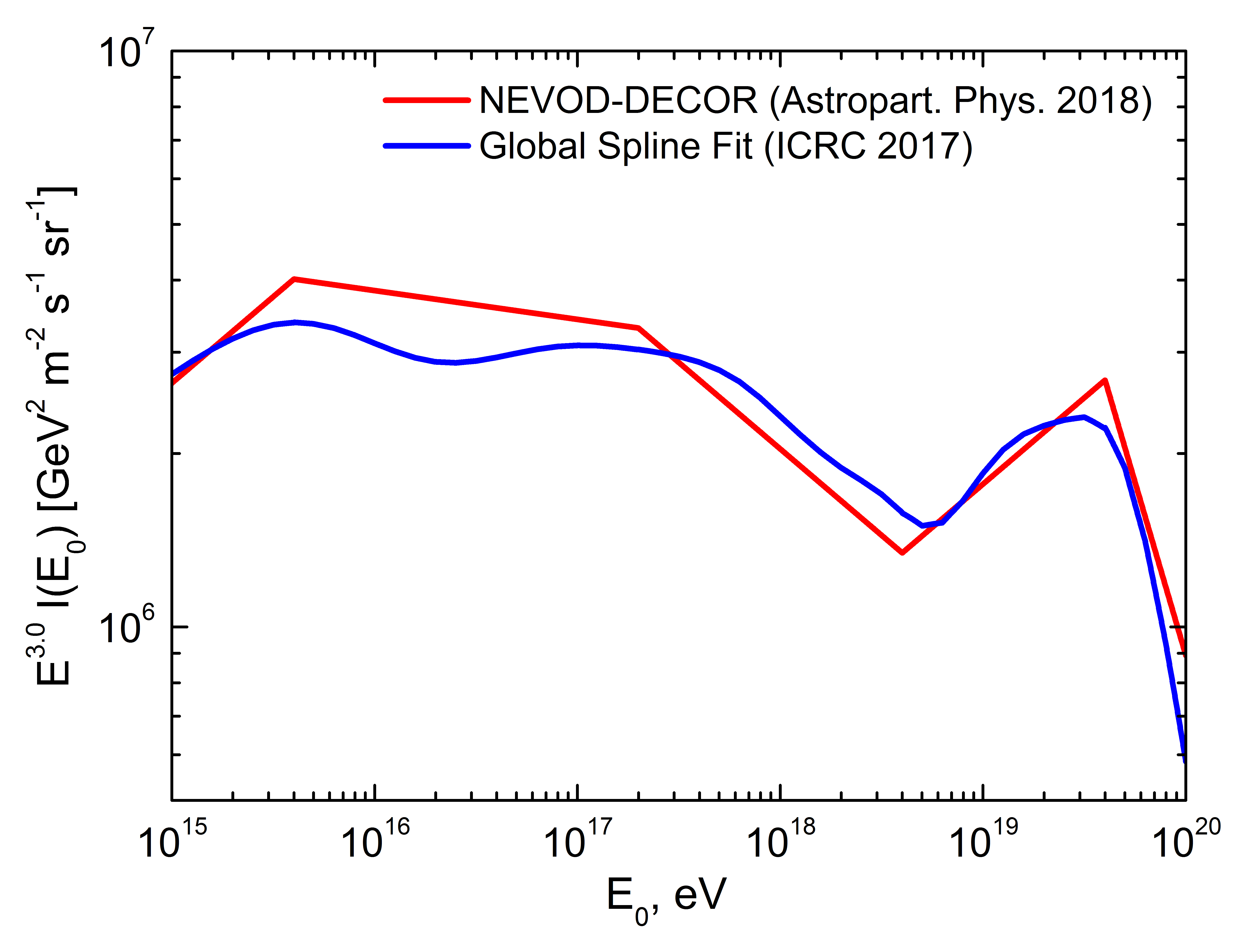}
\caption{PCR energy spectrum models used in [4] (NEVOD-DECOR approximation based on PDG data) and [9] (GSF)}
\label{gsf}
\end{figure}

Generally, the choice of the hadron interaction model and the model of the PCR energy spectrum will affect the estimate of the mass composition of PCRs when interpreting the experimental data. However, the main conclusion that the NEVOD-DECOR data on the intensity of muon bundles can be consistent with calculations only assuming a heavy PCR mass composition in the energy region $\sim$ 10$^{18}$ eV remains.

\section{Energy characteristics of muons in the bundles}
\label{sec:energy}

A drastic contradiction between the experimental data on measuring the density/number of muons in air showers and the $X_{\text{max}}$ by the fluorescence method at ultra-high energies gives reason to believe that a new physical features may emerge, which is not taken into account in modern hadron interaction models. For a deeper understanding of the problem, it is necessary to study the energy characteristics of the EAS muon component, in particular, the muon bundles. Thus, the appearance of an excess of ultra-high energy muons should unambiguously indicate the inclusion of new physical processes (the formation of a new state of matter) \cite{13}.

One of the possible approaches to studying the energy characteristics of the EAS muon component is to measure the energy deposit of muon bundles in the detector material (in our case, in the Cherenkov water calorimeter, see e.g. \cite{14,15,16}), since muon energy loss in the matter almost linearly depends on their energy: $dE/dX \sim a + bE$.

As a measure of the energy deposit of muon bundle in the detector NEVOD, the sum of the all PMTs signals $\Sigma$ (in units of photoelectrons) is used. The assumption that the total yield of Cherenkov light is proportional to the energy losses of muons, including secondary particles and cascades from them, is confirmed by the results of simulating the detector response.

Since, in the first approximation, the total energy deposit $\Sigma$ is proportional to the muon density $D$ in the event (according to the DECOR data), it is more reasonable to consider the specific energy deposit $\Sigma/D$, and we have used this value some time \cite{14,15}.

However, the specific energy deposit depends on the characteristics of the detector. In order to move to traditionally used physical quantities, for example, muon energy losses, it is necessary to find their relationship with the yield of Cherenkov light (signals of the PMTs). For this purpose, a mathematical model of the NEVOD-DECOR setup was developed based on the Geant4 simulation toolkit (v. 10.7) \cite{17}. The calibration of the model was carried out by comparing the results of calculating the response of the NEVOD detector to the passage of near-horizontal muons with experimental data \cite{16}. Near-horizontal muons were sampled using SMs of the detector DECOR located in opposite short galleries of the experimental building. The average energy of such muons is estimated to be about 100 GeV. So, the difference between the average values of the total number of photoelectrons of all PMTs $\Sigma$ in the experiment and simulation is less than 1.5\%, at that the distribution shapes are in good agreement with each other, too. An analysis of the average response of PMTs and QSMs depending on the distance to the muon track, distributions in the number of hit (fired) PMTs and QSMs and other phenomenological characteristics confirmed that the model well (within a few percent) describes the experimental data.

Next, the response of the NEVOD-DECOR setup to artificial muon bundles with a fixed energy of 100 GeV was simulated using a Monte Carlo technique \cite{16}. Both the physical features of the detectors and the conditions for selecting events with muon bundles used in the experiment were taken into account. The distributions of simulated events in muon multiplicity in the bundles (the number of muons that hit the detector), zenith and azimuth angles are almost identical to the experimental ones obtained from the DECOR data.

Special attention was paid to identifying possible systematic distortions that could affect the measurement results. The influence of such factors as residual contribution of the electron-photon and hadron components of the EAS to the response of the detector NEVOD (about 5\% at zenith angles of 55 -- 60$^{\circ}$ and less than 1\% at $\theta =$ 60 -- 65$^{\circ}$), under-estimation of the response due to the digitization threshold, some changes of the registration conditions in different measurement series, efficiency of the DECOR response, masking of tracks, etc. were taken into account.

The desired experimental dependences of the average muon energy in the bundles on the zenith angle and on the local muon density were obtained (see \cite{16} in detail) from the ratio of the measured and calculated specific energy deposits in the NEVOD-DECOR setup. At that, the dependence of the average losses in water on the muon energy normalized to losses at 100 GeV \cite{18} which is practically linear in the region of hundreds of GeV was used.

The expected dependences of the average muon energy in the bundles on the zenith angle and on the local muon density were obtained based on the results of simulations of the EAS muon component (using the CORSIKA program) in a wide range of zenith angles and PCR energies for two extreme (protons and iron nuclei) assumptions about the mass composition and three models of high-energy hadron interactions: QGSJET-II-04, SIBYLL-2.3c and EPOS-LHC. The features of the LMDS method \cite{4} were taken into account: the contribution of the area element $dS$ of the air shower transverse cross section (in the vicinity of the point \textbf{r}) to the flux of detected events is proportional to $\rho(\textbf{r})^{\beta}dS$, where $\rho(\textbf{r})$ is the muon density, $\beta$ is the LMDS slope index. In this case, $\beta = 2.2$ was assumed to be constant; in future, it is supposed to take into account its weak change with increasing PCR energy.

Figure 5 shows for the first time the dependence of the average muon energy in the bundles on the zenith angle. The experimental data of NEVOD-DECOR setup are denoted by symbols; the expected dependences are shown by solid, dashed, and dash-dotted curves for three models of high-energy hadron interactions: QGSJET-II-04, SIBYLL-2.3c, and EPOS-LHC, respectively. The lower curves correspond to calculations for the pure proton PCR mass composition, and the upper ones for iron nuclei. The arrows indicate the typical PCR energies. As can be seen from the figure, there is a growth of the average muon energy in the bundles with an increase in the zenith angle, as well as good agreement between the experiment and theoretical dependences.

Figure 6 shows the dependence of the average muon energy in the bundles on the local muon density (in fact, on the PCR energy) obtained for the zenith angle interval 65 -- 75$^{\circ}$. The symbols show the experimental data; the notations of curves and arrows are the same as in Figure 5. It follows from the figure that the average muon energies in the bundles for the QGSJET-II-04 and SIBYLL-2.3c models only slightly differ from each other, but for the EPOS-LHC model they are approximately 10\% lower. Experimental data indicate an increase in the average muon energy in the bundles at high muon densities corresponding to PCR energies above 10$^{17}$ eV, while the expected muon energies show a tendency to a gradual decrease. The excess of the experiment over the calculations is 3.8$\sigma$ for primary protons and 2.6$\sigma$ for iron nuclei (estimates are given for the QGSJET-II-04 model and the last four experimental points). The deviation of the experimental points from the calculated curves may indicate the inclusion of a new mechanism for the generation of high-energy muons (in the final state) at ultra high energies of PCR. Accumulation of the experimental data and analysis of possible systematic errors are being continued.

\begin{figure}[h]
\begin{minipage}[h]{0.49\linewidth}
\centering
\includegraphics[width=1.\linewidth]{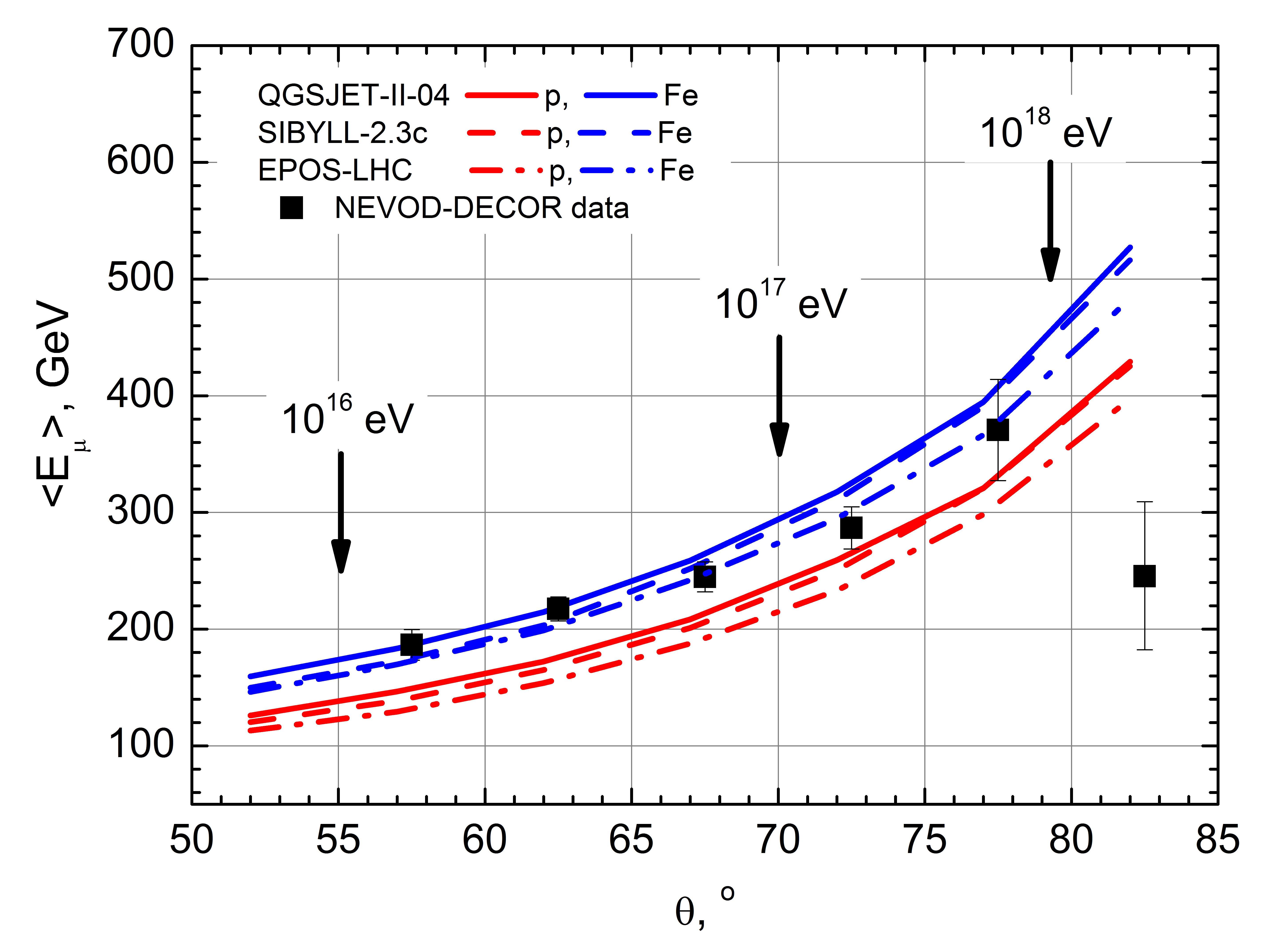}
\captionsetup{width=0.95\linewidth}
\caption{Dependences of the average energy of muons in the bundles on the zenith angle}
\end{minipage}
\hfill
\begin{minipage}[h]{0.49\linewidth}
\centering
\includegraphics[width=1.\linewidth]{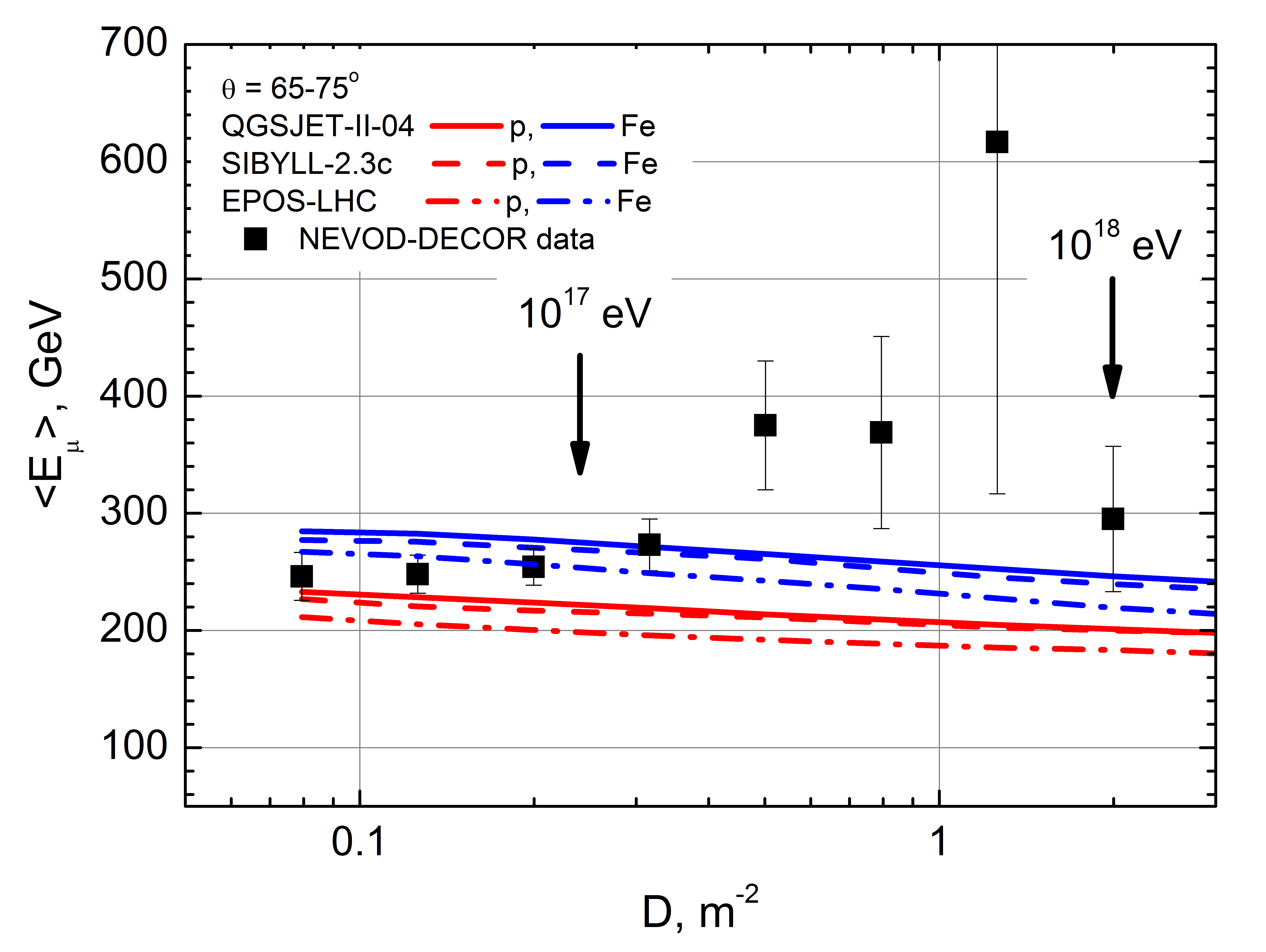}
\captionsetup{width=0.95\linewidth}
\caption{Dependences of the average energy of muons in the bundles on the local muon density}
\end{minipage}
\label{energy}
\end{figure}

\section{Conclusions}
The present data on the intensity of the muon bundles observed in the NEVOD-DECOR experiment are only compatible with calculations if we assume extremely heavy (iron group nuclei) mass composition at the primary energies around $10^{18}$ eV. However, such an assumption contradicts the available fluorescence data on $X_{\text{max}}$ which favor a light (proton, He) mass composition at these energies. It seems that the solution of the ``muon puzzle'' requires the introduction of serious changes in existing hadron interaction models, which could change the balance between muon and electron-photon EAS components.

For the first time, the dependences of the average energy of muons in the bundles on the zenith angle and local density in the region corresponding to the energies of primary particles 10 -- 1000 PeV have been obtained. An increase in the average energy of muons in the bundles is observed in comparison with the expected one for primary energies above $10^{17}$ eV. This may indicate the inclusion of a new mechanism for the generation of high-energy muons.

\section*{Acknowledgements}
The work was performed at the Unique Scientific Facility ``Experimental complex NEVOD'' with the financial support provided by the Russian Ministry of Science and Higher Education, project ``Fundamental problems of cosmic rays and dark matter'', No. 0723-2020-0040. Simulations were carried out using the resources of the MEPhI high-performance computing center.

\nolinenumbers

\end{document}